# Polymorphism of monatomic iodine

Alexander F. Goncharov[1], Huawei Chen[1,2], Iskander G. Batyrev[3], Maxim Bykov[4], Lukas Brüning[4], Elena Bykova[5], Valentin Kovalev[5], Mohammad F. Mahmood[2], Mohamed Mezouar[6], Gaston Garbarino[6], Jonathan Wright[6]

[1] Earth and Planets Laboratory, Carnegie Science, Washington, DC 20015, USA
[2] Department of Mathematics, Howard University, Washington DC 20059 USA.
[3] U.S. Army Research Laboratory, RDRLWML-B, Aberdeen Proving Ground, Maryland 21005, United States
[4] Institute of Inorganic and Analytical Chemistry, Goethe University Frankfurt, Max-von-Laue-Straße 7, 60438 Frankfurt am Main, Germany
[5] Institute of Geosciences, Goethe University Frankfurt, Altenhöferallee 1, 60438 Frankfurt am Main, Germany
[6] European Synchrotron Radiation Facility BP 220, 38043 Grenoble Cedex, France

**We applied synchrotron single-crystal X-ray diffraction in a diamond anvil cell at 48-51 GPa and first-principles theoretical calculations to study the crystal structure of solid atomic iodine at high pressure. We report the synthesis of two phases of atomic iodine at 48-51 GPa via laser heating of I-N$_2$ mixtures. Unlike the familiar monatomic *I*4/*mmm* structure, which consists of crystallographically equivalent atoms, a new *Pm$\bar{3}$n* structure is of inclusion type, featuring two distinct kinds of atoms: a central detached one and peripheral ones forming the linear chains. Moreover, we observe crystallization of the familiar high-pressure face centered cubic (fcc) structure, albeit at much lower pressures compared to cold compressed iodine. The discovery of *Pm$\bar{3}$n* structure in iodine marks an important step in understanding of the pressure induced phase transition sequence in halogens.**

Applying high pressure to pnictogens, chalcogens, and halogens (including hydrogen), results in molecular dissociation and formation of monatomic solids. Investigation of such transitions greatly contributes to our understanding of the evolution of chemical bonding at high pressure. Monatomic high-pressure solids are of interest due to their relevance to the insulator-metal transition, high-temperature superconductivity (1), high energy-density materials (2), and other phenomena of interest for fundamental science and applications. As the largest radii element among the diatomic molecular solids of the halogen groups, iodine experiences molecular dissociation at much lower pressure, making it feasible for experiments aiming at understanding the progressive formation of monoatomic state in diatomic molecules (3).

Halogens (X) at near to ambient conditions represent a typical case of diatomic molecules, crystallized in molecular solids with a herringbone-like structure such as the phase I (space group *Cmca*) for Cl$_2$, Br$_2$, and I$_2$. These solids experience a series of phase transformations on their pathway to metallization and molecular dissociation. The structural sequence includes phases with mixed atomic associations, such as elongated X$_2$ and linear X$_3$ atomic units forming complex incommensurate structures (3-5). On the other hand, monatomic phases appeared to be less complex typical for simple metals.

The sequence of phase transformation in monatomic iodine is as follows: *Immm* (Phase II) – (43 GPa) - *I4/mmm* (Phase III) – (55 GPa with 4 GPa hysteresis upon reversal), and *Fm$\bar{3}$m* (Phase IV, fcc) (stable up to 276 GPa (6)) (7). In lighter halogens, the transitions occur at higher pressures, and moreover, the phase sequence can be, in principle, different. In chlorine, the phase sequence seems to be similar to iodine, the transition to *Immm* Phase-II is reported experimentally at 266 GPa (8). Theory predicts this transition at 133 GPa with a further transition to an fcc structure above 330 GPa (9). Other theory (10) suggests these transitions to occur at higher pressures, 157 and 372 GPa, respectively. In contrast, molecular fluorine $F_2$ is predicted to behave differently, transforming from a familiar molecular *Cmca* phase to a tetragonal monatomic tetragonal *P4$_2$/mmc* and then to a cubic *Pm$\bar{3}$n* structure at 2500 and 3000 GPa, respectively (9). These two metallic phases differ from common simple metallic structures in that there are two distinct atomic positions suggesting an unusual chemical bonding. Here, we report the observation of a cubic *Pm$\bar{3}$n* structure of iodine in the laser heating experiments at 48-51 GPa, where molecular nitrogen served as a pressure medium and potential reagent. Moreover, we find that other iodine grains form an fcc structure, which was reported to become stable at substantially higher pressures.

We performed the experiments in diamond anvil cells (DAC) designed for single-crystal X-ray diffraction. Similar to our previous work on iodine (3), small pieces of single-crystal $I_2$ were loaded in a cavity of a DAC with culets of 200 μm in diameter made in a pre-indented rhenium gasket. The sample was loaded in compressed to 150 MPa $N_2$ gas of high purity. The Raman peak positions of solid $N_2$ were used to determine pressure in the DAC (11). The samples were compressed to 48-52 GPa and laser heated to 2000-2500 K.

For single-crystal X-ray diffraction (XRD) investigation at ESRF ID27 (12) and ID11, we used a monochromatic X-ray beam of <3 μm in diameter with the wavelength of 0.3738 Å (ID27) and 0.2846 Å (ID11), respectively. The procedure of single-crystal XRD and the structure determination has been described in the previous publications (e.g., Refs. (3, 13)).

First-principles theoretical calculations have been performed in the phases of interest at selected pressures. The structures of these phases were optimized utilizing norm-conserving pseudopotentials using the generalized gradient approximations (GGA)-Perdew-BurkeErnzerhof (PBE) functional with the D3(BJ) dispersion correction. Monkhorst-Pack grid size for *k*-point sampling of the Brillouin zone was 6×7×6 (14). The electronic band-gap calculations have been performed within GGA/PBE and PBE-based screen hybrid functional Heyd-Scuseria-Ernzerhof (HSE06) (15) approximations.

Laser heating of iodine at 48-52 GPa above 2000 K results in the sample recrystallization in the hot spot. At 48 GPa, no reaction between iodine and $N_2$ has been detected. The XRD is presented by a mixture of iodine phases and molecular nitrogen ε-$N_2$ (16). Iodine primarily crystallizes in the fcc structure (iodine IV (7)), well documented in powder (Fig. 1) and single-crystal diffraction patterns (Fig. S1 of Supplementary Materials). Additionally, another iodine phase with a *Pm$\bar{3}$n* structure is identified alongside the main phase, referred to as iodine-VII hereafter. The structure was solved using single-crystal XRD data at 48 GPa (Fig. 2) with the SHELXT structure solution program (17) and refined with the OLEX2 program (18, 19) (Table SI). The Cambridge Structural Database contain the supplementary crystallographic data for this work. These data can be obtained free of charge from FIZ Karlsruhe (deposition numbers CSD 2380946) (20). The iodine VII phase was also detected in powder diffraction patterns (Fig. 1); the lattice parameters refined from these measurements yielded the values consistent with the results of SC XRD. The common *Immm* and *I4/mmm* phases (iodine II and iodine III, respectively, (7)) previously reported at these pressure conditions are not observed in laser heated samples; nonetheless, we observed *Immm* phase in a separate powder diffraction experiment at 52 GPa in a sample, which was compressed in $N_2$ medium at room temperature (Fig. S2 of Supplementary Materials).

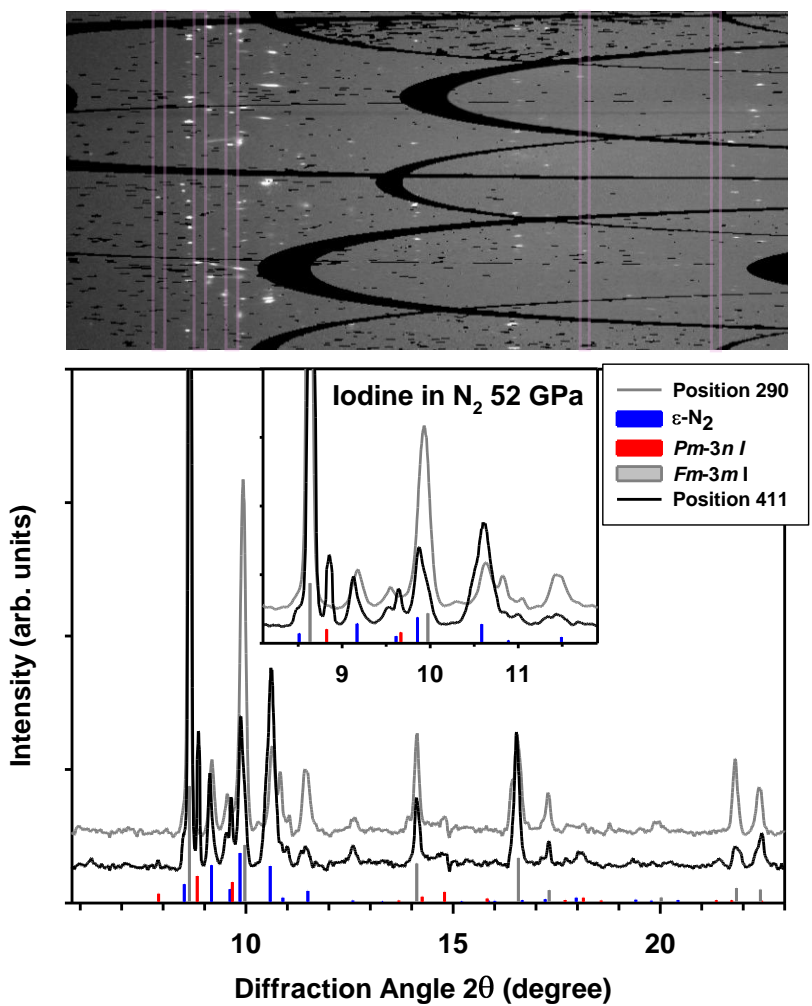

**Figure 1.** X-ray diffraction patterns of I-$N_2$ samples after laser heating at 51 GPa. Both integrated curves are measured away from the heating spot. Inset shows an expanded view near the most characteristic peaks. Vertical bars correspond to Bragg peaks of the fitted phases. The top curve (shifted for clarity) can be well represented by a mixture of diffraction patterns of molecular ε-$N_2$ and fcc iodine. The bottom curve reveals the peaks of $Pm\bar{3}n$ iodine in addition to those of ε-$N_2$ and fcc iodine. The top panel is a 2D image (position 411, where $Pm\bar{3}n$ iodine is present) in rectangular coordinates. Light red rectangles box single-crystal-like diffraction peaks of $Pm\bar{3}n$ iodine. The X-ray wavelength is 0.3738 Å.

Our experiments demonstrate that two additional phases of monatomic iodine can be crystallized at approximately 50 GPa in the laser annealed samples (Fig. 3). Although these two phases have similar atomic volumes (Fig. 4), the crystal structures are different, which results in very diverse nearest neighbor interatomic distances (Fig. 3). In $Pm\bar{3}n$ iodine-VII, three quarters of iodine atoms form chains along all 3 crystallographic directions, while the rest of the atoms occupies the symmetric crystallographic cites, which are much less bound to other atoms; this makes iodine VII structure of a host-guest type. The fcc iodine is a high-pressure phase, which forms upon compression above 55 GPa at room temperature (7). The crystallization of this phase at lower pressure in this work, is likely the results of the temperature annealing promoting the formation of a more thermodynamically stable phase. The fcc iodine is denser than the $Immm$ and $Pm\bar{3}n$ iodine modifications (Fig. 4). This is consistent with the observations of Ref. (7) reporting the volume drop at the transitions to fcc iodine IV.

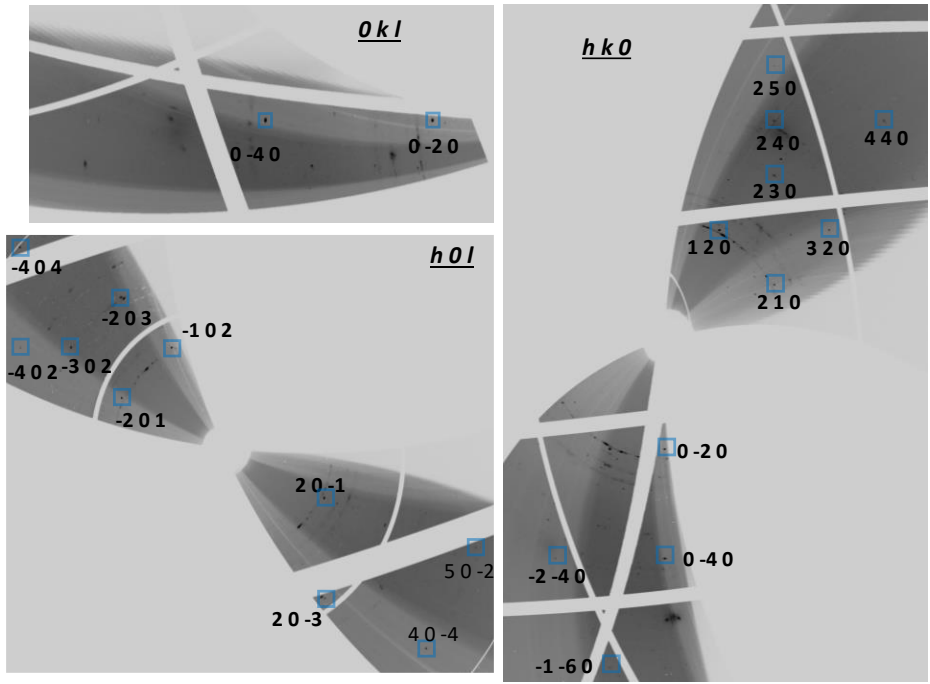

**Figure 2.** Reconstructed reciprocal lattice planes of $Pm\bar{3}n$ iodine at 48 GPa. Observed Bragg diffraction spots from sample marked by blue rectangles have been indexed and used to determine the structure (see details in Table S1). For $Pm\bar{3}n$ iodine VII with the atoms occupying the Wyckoff sites $2a$ and $6d$, the reflection conditions are $h + k + l = 2n$ and $h = 2n + 1$ and $k = 4n$, $l = 4n + 2$ ($h,k,l$ permutable), respectively. These conditions are consistent with the observations.

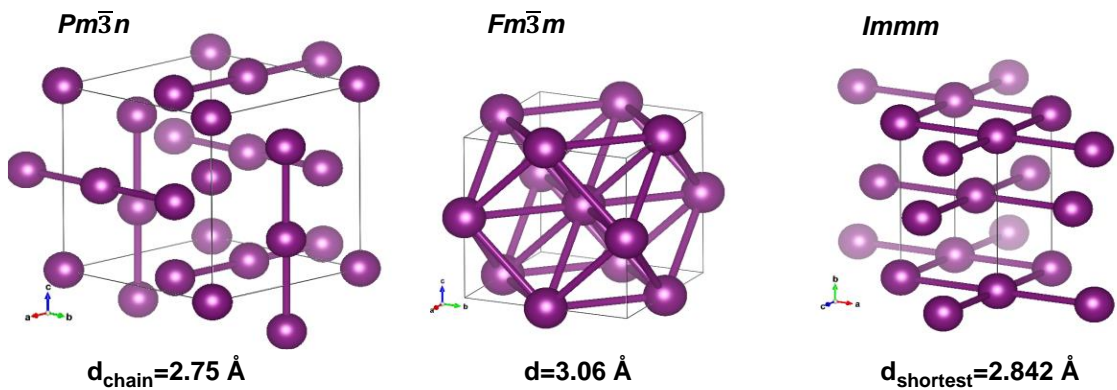

**Figure 3.** Crystal structures of iodine determined here at ≈50 GPa. The shortest nearest-neighbor interatomic distances are shown by bars and their values are presented.

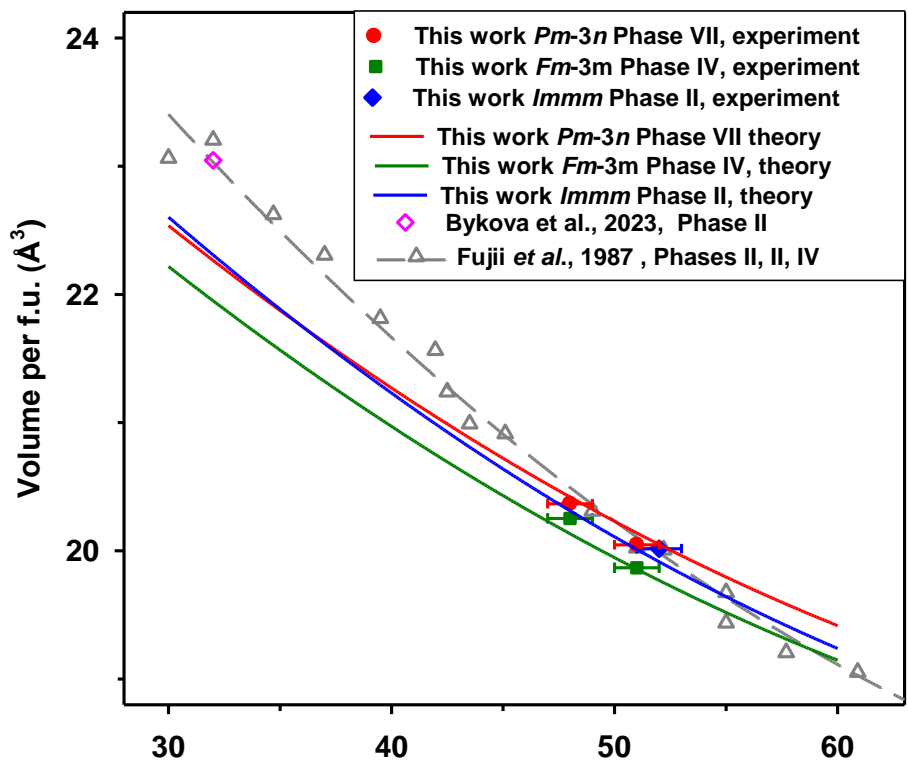

**Figure 4.** The atomic volume of high-pressure monatomic phases of iodine. Filled symbols show the experimental results of this work (the uncertainty is smaller than the size of the symbol). Open symbols are the results of previous works: triangles- the volumes of iodine-II, III, and IV, the dashed line is a guide to the eye (7); diamond – the volume of iodine-II (3). Solid lines are the theoretical calculation of this work; the result for $I4/mmm$ phase (iodine-III) is not shown for clarity as it is very close to that for $Immm$ phase II.

A new $Pm\bar{3}n$ iodine is noticeably less dense than fcc $Fm\bar{3}m$ iodine (Fig. 4) even though both phases have been crystallized at the same P-T conditions. The comparison of theoretically computed enthalpies (Fig. 5(a)) shows that $Fm\bar{3}m$ iodine becomes energetically favorable with respect to the common *I4/mmm* phase (which is close in enthalpy to *Immm* phase, as the structure of the latter one is a mere distortion of the *I4/mmm* structure) above 50 GPa, while new $Pm\bar{3}n$ iodine-VII is metastable at 30-60 GPa. When considering the temperature dependent contribution (see Supplementary notes) to compute the free energies (Fig. 5), then $Fm\bar{3}m$ iodine becomes stable at 47 GPa while $Pm\bar{3}n$ iodine-VII becomes closer to the stability at 300 K. $Fm\bar{3}m$ iodine is thermodynamically stable at the experimental pressure of ≈50 GPa, but the difference in the Gibbs free energy with the new $Pm\bar{3}n$ iodine phase is only about 20 meV at 300 K. The difference in the Gibbs free energy between the $Pm\bar{3}n$ iodine and *I4/mmm* phase reduces with temperature and vanishes at 1500 K due to the larger entropy contribution of $Pm\bar{3}n$ iodine (Fig. S3 of Supplementary Materials). These considerations explain the experimental observations of $Pm\bar{3}n$ iodine at high temperature. The experiments show that $Fm\bar{3}m$ is the major phase after laser annealing at 48 GPa (Fig. 1), while $Pm\bar{3}n$ iodine is kinetically stable due to a drastic difference in structure, which causes a kinetic barrier for the transformation to thermodynamically stable $Fm\bar{3}m$ iodine.

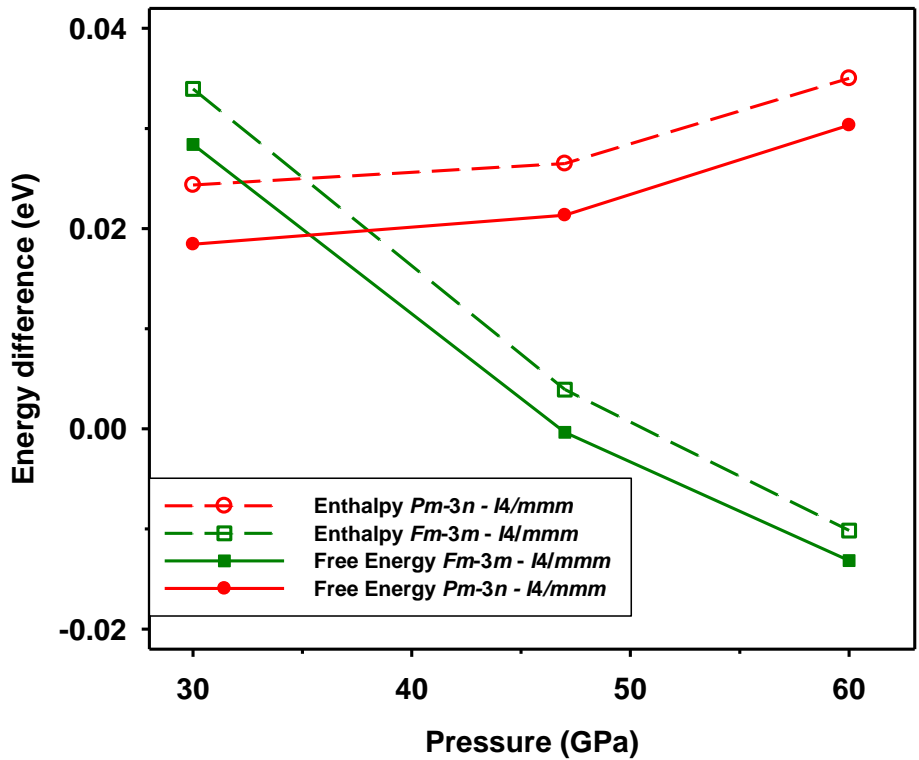

**Figure 5.** Theoretically computed here from first principles the enthalpies and Gibbs free energies vs pressure of the phases of interest plotted with respect to results for *I4/mmm* phase.

The electronic structure of $Pm\bar{3}n$ phase-VII, which has 2 different site symmetries for iodine atoms, is peculiar compared to that of fcc $Fm\bar{3}m$ iodine. Our calculations (Fig. S4 of Supplementary Materials) show that both phases are metallic with the comparable values of the density of states at the Fermi level. Also, we calculated the electron localization functions for $Pm\bar{3}n$ iodine-VII at several pressures (Fig. 6). They show that there is a substantial electronic density between the atoms in the chains at 47 GPa, which corresponds to the formation of chemical bonds in these directions. We speculate that the presence of such bonds provides an electronic mechanism of structural stabilization (Peierls distortion) of long-chain $Pm\bar{3}n$ iodine-VII. Our phonon dispersion calculations show that there are no imaginary frequencies suggesting that $Pm\bar{3}n$ iodine-VII is dynamically stable (Fig. S5 of Supplementary Materials).

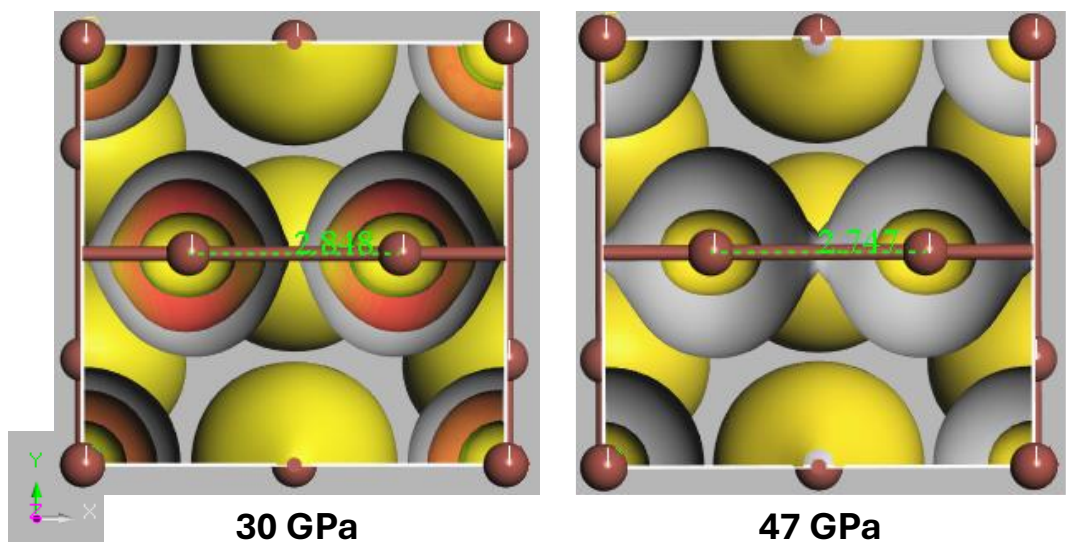

**30 GPa**          **47 GPa**

**Figure 6.** Theoretically computed Electron Localization Functions at ELF=0.55 of $Pm\bar{3}n$ iodine phase-VII at 30 and 47 GPa.

Overall, our work shows that the predicted previously for chlorine above 3000 GPa $Pm\bar{3}n$ phase (9) is realized at some 50 GPa in iodine. The (meta)stability of this phase is entropy driven due to a large entropy contribution, related to the presence of indefinitely long strongly bound atomic chains. This result adds to our understanding of polymorphism of halogens at high pressures and call for investigations of other halogens (e.g., bromine).

Support is acknowledged from the National Science Foundation DMR-2200670, CHE-2302437, and Carnegie Science. M.B. acknowledges the support of Deutsche Forschungsgemeinschaft (DFG Emmy-Noether project BY112/2-1). E.B. acknowledges financial support from the program 'Promotion of Equal Opportunities for Women in Research and Teaching' funded by the Free State of Bavaria.

Supplementary materials to

# Polymorphism of monatomic iodine

Alexander F. Goncharov[1], Huawei Chen[1,2], Iskander G. Batyrev[3], Maxim Bykov[4], Lukas Brüning[4], Elena Bykova[5], Valentin Kovalev[5], Mohammad F. Mahmood[2], Mohamed Mezouar[6], Gaston Garbarino[6], Jonathan Wright[6]

[1] Earth and Planets Laboratory, Carnegie Science, Washington, DC 20015, USA
[2] Department of Mathematics, Howard University, Washington DC 20059 USA.
[3] U.S. Army Research Laboratory, RDRLWML-B, Aberdeen Proving Ground, Maryland 21005, United States
[4] Institute of Inorganic and Analytical Chemistry, Goethe University Frankfurt, Max-von-Laue-Straße 7, 60438 Frankfurt am Main, Germany
[5] Institute of Geosciences, Goethe University Frankfurt, Altenhöferallee 1, 60438 Frankfurt am Main, Germany
[6] European Synchrotron Radiation Facility BP 220, 38043 Grenoble Cedex, France

Entropy contribution to the Gibbs Free Energy

Zero-point energy:

ZPE=h/2*$\int g(w)w\,dw$, where g(w) – phonon density found at 0 K.

$\Delta G$= ZPE + kT$\int g(w) \ln\left[1 - \exp\left(-\frac{hw}{kT}\right)\right] dw$

it includes the entropy contribution.

The Gibbs free energy:

G= Eo+ $\Delta$G,

Where Eo is the enthalpy.

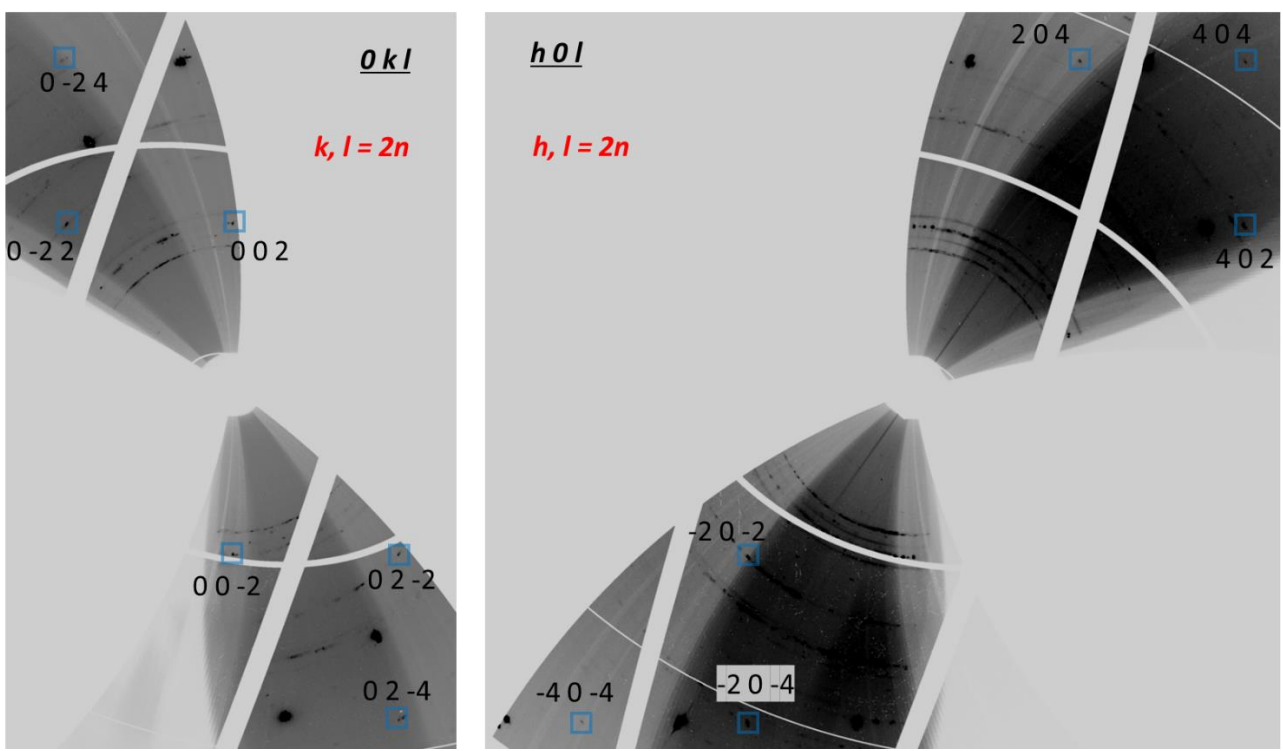

**Figure S1.** Reconstructed reciprocal lattice planes of $Fm\overline{3}m$ iodine at 49 GPa. Observed Bragg diffraction spots from sample marked by blue rectangles have been indexed and used to determine the structure. The X-ray extinctions rule for this phase is depicted.

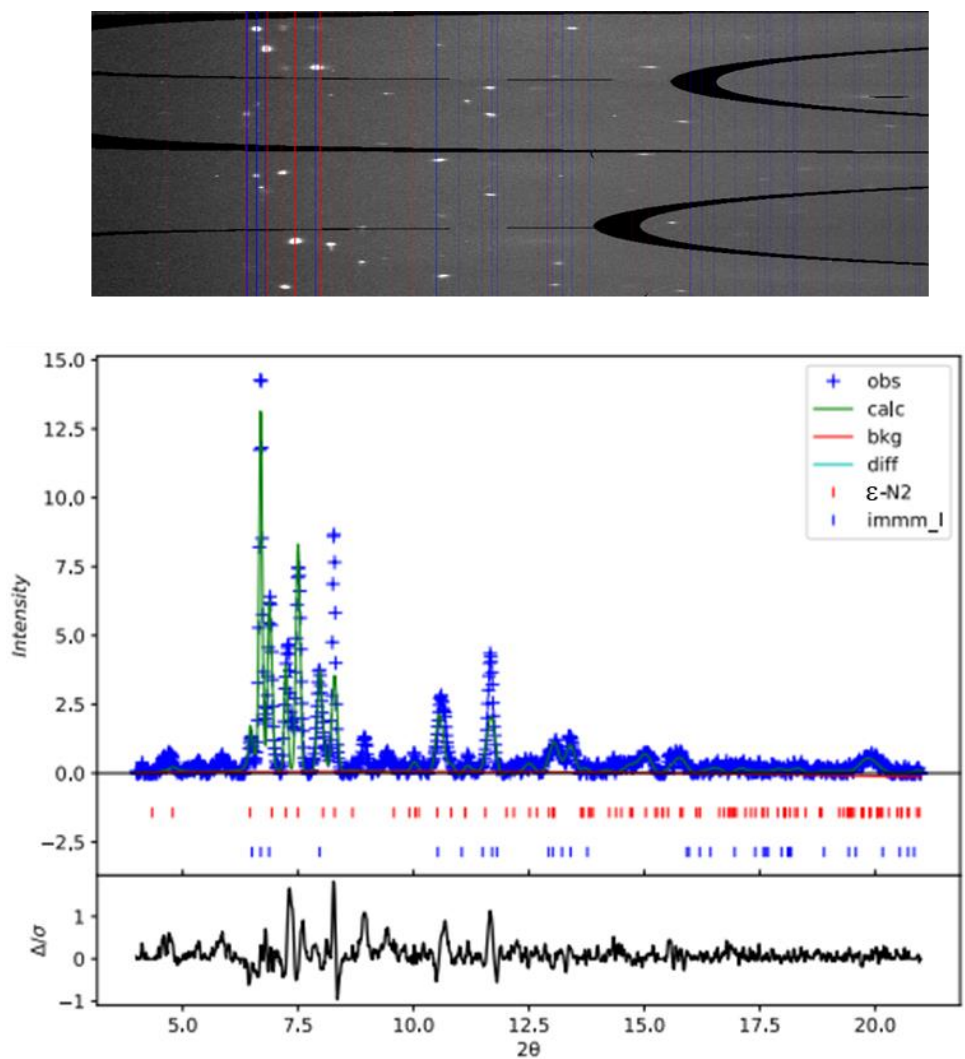

**Figure S2.** XRD pattern of iodine in $N_2$ medium at 52 GPa. In the main panel, crosses are 1D integrated data, green line is the Rietveld refinement assuming that there are two phases: *Immm* and $\varepsilon$-$N_2$. The refined lattice parameters are $a$=2.9698(5) Å, $b$= 4.743(8) Å, $c$=2.842(3) Å for *Immm* Iodine-II and a=6.8239(2) Å, c=9.7429(3) Å for $\varepsilon$-$N_2$. The top panel is a 2D XRD pattern in rectangular coordinates, the positions of the major phases are marked by vertical lines. The X-ray wavelength is 0.2846 Å.

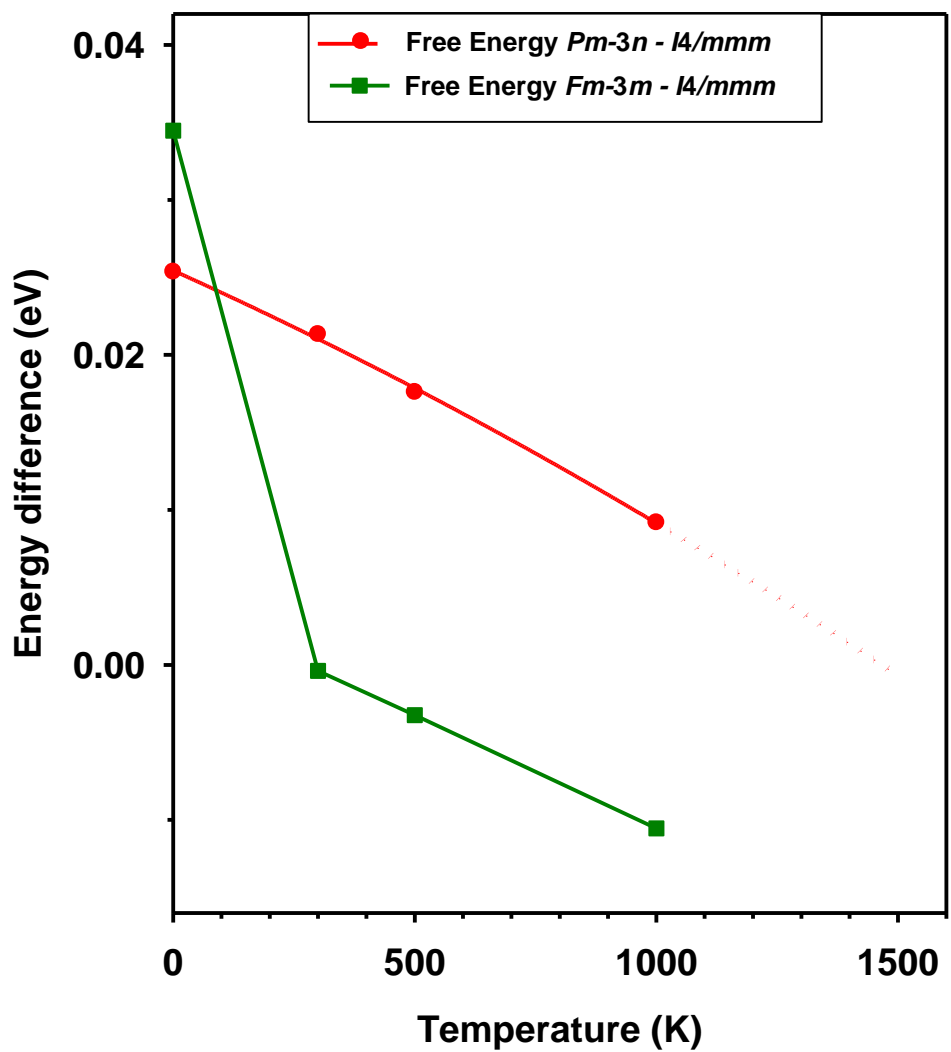

**Figure S3.** Theoretically computed here from first-principles the Gibbs free energies vs temperature of the phases of interest plotted with respect to results for *I4/mmm* phase.

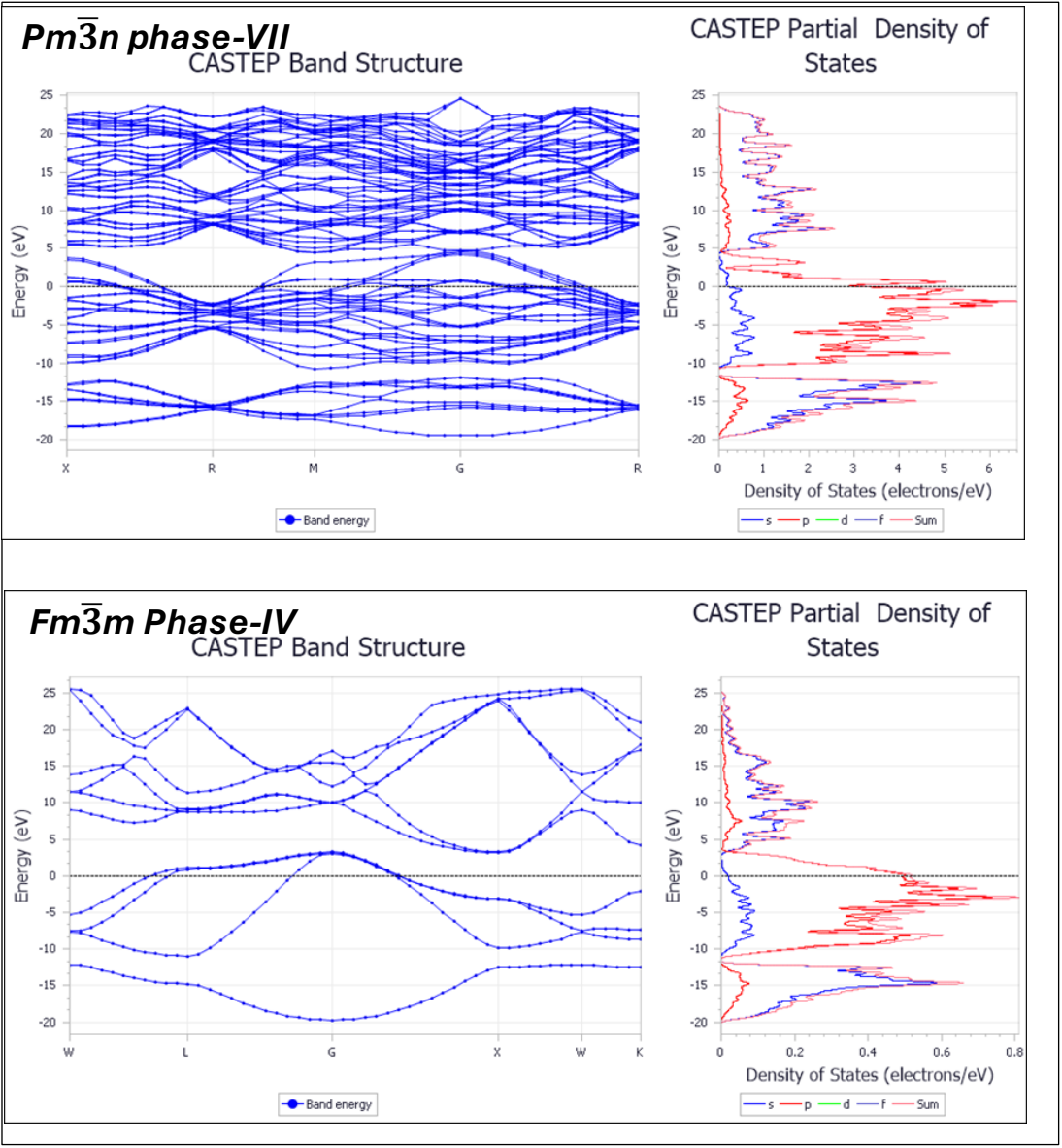

**Figure S4.** The calculated electronic band structure and electronic density of states of $Pm\bar{3}n$ iodine-VII in comparison to those of $Fm\bar{3}m$ iodine-IV at 47 GPa. The electronic density of states should be divided by number of atoms in the unit cell, which makes the density of states at the Fermi level about the same for $Pm\bar{3}n$ than for $Fm\bar{3}m$ phase.

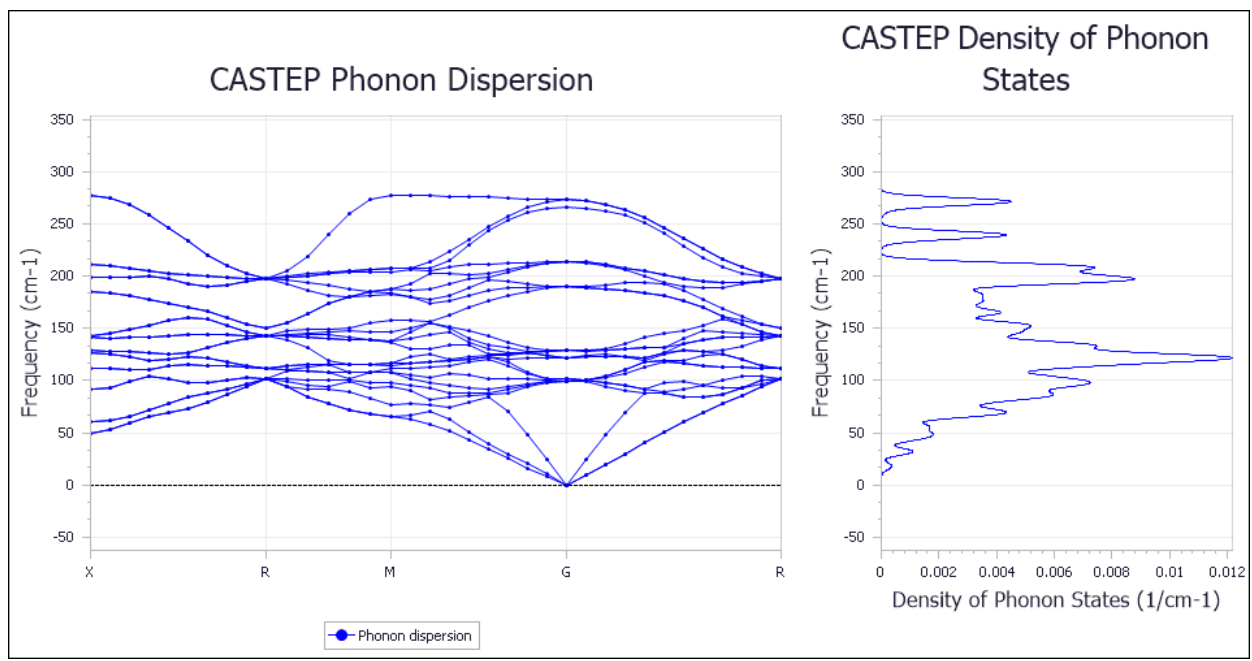

**Figure S5.** The calculated phonon dispersion structure and phonon density of states of $Pm\bar{3}n$ iodine-VII at 47 GPa.

**Table S1. Details of crystal structure refinements for $Pm\bar{3}n$ iodine at high pressures**

| Pressure | 47 GPa |
|---|---|
| Crystal data | |
| Chemical formula | I |
| $M_r$ | 126.90 |
| Crystal system, space group | Cubic, $Pm\bar{3}n$ |
| Temperature (K) | 293 |
| $a$ (Å) | 5.4618 (17) |
| $V$ (Å$^3$) | 162.93 (15) |
| $Z$ | 8 |
| Radiation type | Synchrotron, $\lambda$ = 0.3738 Å; ID27 (ESRF) |
| $\mu$ (mm$^{-1}$) | 6.78 |
| Crystal size (μm) | 1 |
| | |
| Data collection | |
| Diffractometer | Esperanto-*CrysAlis PRO*-abstract goniometer imported esperanto images |
| Absorption correction | Multi-scan *CrysAlis PRO* 1.171.43.95a (Rigaku Oxford Diffraction, 2023) Empirical absorption correction using spherical harmonics, implemented in SCALE3 ABSPACK scaling algorithm. |
| $T_{min}$, $T_{max}$ | 0.674, 1.000 |
| No. of measured, independent and observed [$I > 2\sigma(I)$] reflections | 312, 78, 39 |
| $R_{int}$ | 0.092 |
| $(\sin\theta/\lambda)_{max}$ (Å$^{-1}$) | 0.920 |
| Atomic positions I001 I 0.500000 1.000000 0.750000 0.0172(9) Uani 1 8 d S T P . . I002 I 0.500000 0.500000 0.500000 0.0188(13) Uani 1 24 d S T P . . | |
| Refinement | |
| $R[F^2 > 2\sigma(F^2)]$, $wR(F^2)$, $S$ | 0.061, 0.171, 0.90 |
| No. of reflections | 78 |
| No. of parameters | 4 |
| $\Delta\rangle_{max}$, $\Delta\rangle_{min}$ (e Å$^{-3}$) | 2.41, -3.54 |

Computer programs: *CrysAlis PRO* 1.171.43.95a (Rigaku OD, 2023) [1],

SHELXT 2018/2 [2], olex2.refine 1.5-ac5-024 [3], *SHELXL2018*/3 [2], Olex2 1.5-ac5-024 [4].